\documentstyle[12pt]{article}

\input epsf

\newcommand{\nc}{\newcommand}
\nc{\sh}[1]{\!\not\! #1}
\nc{\ga}[1]{\gamma_{#1}}
\nc{\g}{\gamma}
\nc{\gm}{\gamma_{\mu}}
\nc{\gf}{\gamma_{5}}
\nc{\fiv}[3]{\phi_{#1}^{#2} (v_{#3})}
\nc{\f}[2]{\phi_{#1}^{#2}}
\nc{\F}[2]{\Phi_{#1}^{#2}}
\nc{\fm}[3]{\phi_{\mu #1}^{\;\; #2} ({#3})}
\nc{\fn}[3]{\phi_{\nu #1}^{\;\; #2} ({#3})}
\nc{\ff}[3]{\phi_{5 #1}^{\;\; #2} ({#3})}
\nc{\Tr}{ \, {\mbox {\rm Tr}}[}
\nc{\Li}{{\cal L}_{int}}
\nc{\decay}[2]{\Gamma_{#1 \rightarrow #2}}
\nc{\lam}[1]{\lambda^{#1}(m_{1}^{2},m_{2}^{2},m_{3}^{2})}
\nc{\mod}[1]{\mid\! #1 \!\mid}
\nc{\mo}{m_{1}}
\nc{\mtw}{m_{2}}
\nc{\mth}{m_{3}}
\nc{\po}{p_{1}}
\nc{\ptw}{p_{2}}
\nc{\pth}{p_{3}}
\nc{\dg}[1]{#1^{\circ}}
\nc{\gvpp}{g_{_{VPP}}}
\nc{\gvppd}{g_{_{VPP}}^{\prime}}
\nc{\gvvp}{g_{_{VVP}}}
\nc{\gvpg}{g_{_{VP}\g}}
\nc{\gvvv}{g_{_{VVV}}}
\nc{\tV}{\theta_{V}}
\nc{\tP}{\theta_{P}}

\nc{\sm}[1]{\small #1 \normalsize}
\nc{\ty}[1]{\tiny #1 \normalsize}

\nc{\beq}{\begin{equation}}
\nc{\eeq}{\end{equation}}
\nc{\bea}{\begin{eqnarray}}
\nc{\eea}{\end{eqnarray}}
\nc{\beas}{\begin{eqnarray*}}
\nc{\eeas}{\end{eqnarray*}}
\nc{\tfrac}[2]{{\textstyle \frac{#1}{#2}}}

\begin{document}
\hfill UTAS--PHYS--94-26

\vspace{1in}

\Large
\begin{center}
{\bf Meson Supermultiplet Decay Constants}
\large
\vspace{.7in}

{\em N.R. Jones and R. Delbourgo}
\vspace{.1in}

\normalsize
Physics Department, University of Tasmania,\\
 GPO Box 252C, Hobart, Tas. 7001, Australia.
\vspace{1in}
\end{center}

\begin{abstract}
We use a covariant supermultiplet theory to determine the primary coupling 
constant associated with several types of two-body meson decay.  Despite the
diverse range of decays considered the primary coupling constant is surprisingly
uniform.  We envisage the extension of the  techniques to heavy quark 
cases, including as preliminary examples the calculation of the
$D^{*+}$ and $D^{*0}$ total decay widths with results $57.7 \pm 1.5$ KeV and
$42.5 \pm 2.6$ KeV respectively, as well as some predictions about $D^*$ and 
$B^*$ radiative decays.
\end{abstract}

\newpage
\normalsize
\section{Introduction}

In these heady days of heavy quark effective theories (HQET) we feel it is
appropriate to review a supermultiplet scheme developed in the mid-1960's which
has great similarities to the HQET in respect of the heavy quark and the accompanying 
`brown muck'.  In doing so we hope to test
the covariant supermultiplet theory for light as well as heavy degrees of freedom,
 assessing the extent of symmetry breaking and how it manifests itself.
We may then apply our techniques to heavy quark examples in future work with 
considerable confidence.  

To determine the matrix elements of currents between hadrons requires
knowledge of the hadronic wave function in terms of the quark and gluon
constituents.  A full relativistic treatment of such constituents is impossible
because of the infinite degrees of freedom associated with the quarks and
gluons.  However, the success of the non-relativistic quark model spawned
several workers \cite{delb65a,beg65,saki65} to construct relativistic spinor
fields describing pointlike mesons incorporating the correct spin, parity,
flavour and colour degrees of freedom.  Despite later evidence that the mesons
were not pointlike objects, these group theoretical approaches can be shown to
be equivalent to the non-relativistic weak binding limit \cite{huss91b}.

The basis of our work is a relativistic meson supermultiplet field
\cite{salam65}.  The wavefunction describing the meson is dynamically 
equivalent to a system of two quarks, both of which are on-shell and moving at
the same velocity. This differs little from the heavy quark picture which 
assumes that a meson with a quark much heavier than its light antiparticle partner
will have the heavy component almost on-shell and moving at the same velocity
as the meson because the light `brown muck' \cite{flyn} must move with the same velocity.
  This assumption, along with the hope that weak interactions will
not affect the motion of the heavy quark at small recoil, led to the now popular
 heavy quark symmetry and decoupling (see \cite{mann93,grin92} for a review). 

The reasons for using the supermultiplet scheme are several.  Firstly, the
scheme automatically incorporates the Zweig rules and duality diagrams, so  
one can easily determine
the Zweig allowed strong decays.  At the same time it incorporates isospin
and field mixing factors so that one can readily normalise the coupling
constant of various decays; indeed this makes supermultiplet theory very predictive
as one need only know a single coupling constant to predict
widths of many seemingly unrelated processes.  Secondly, radiative decay modes
may be examined by combining the supermultiplet scheme with the vector meson
dominance model. This permits the theory to make some quite accurate
predictions about photon mediated decays such as $\omega \to \pi^+ \pi^-$.
Thirdly, excited mesonic states may be constructed in terms of the supermultiplet
field \cite{delb94} so we are able to further broaden its applications.  Finally, 
the scheme is easily extended to include the $c$ \cite{delb75} and $b$ quark 
flavoured mesons so that we may venture into the heavy quark arena with little 
modification.

The recent ACCMOR \cite{accm92} and CLEO Collaborations \cite{cleo92a}
 have renewed interest in the
$D^*$ decays due to two major findings.  The $D^{*+}$ total width was
measured with an upper bound of 131 KeV (signficantly lower than the 1992 
upper bound of 1.1 MeV), and the $D^{*+} \to D^+ \g$ branching fraction
appears significantly smaller than earlier measurements.  Both
these findings are consistent with constituent quark model \cite{kama92} and 
HQET predictions \cite{cola9307}.  We provide similar calculations within the
supermultiplet framework and reproduce these findings with considerably far 
ease. 

\section{Supermultiplet Field}

Derivation of the wave function \cite{delb65a} proceeds by applying a
relativistic boost to the rest frame spinor $\f{a}{b}(\hat{p})$, odd under parity.
This leads to a relativistic spinor $\f{A}{B}(p)$, satisfying Bargmann--Wigner equations, namely
\beq
 \f{A}{B}(p) = \f{a \alpha}{b \beta} (p) = 
  (\sh{p}+m) [\gamma^{\mu} \f{\mu a}{\;\; b} - \gf \f{5 a}{\;\; b}]_{\alpha}^{\beta} /2m
  \label{eq:superfield}
\eeq
where $a,b$ are flavour indices, $\alpha,\beta$ are spin indices and $\f{5}{}$
corresponds to the pseudoscalar nonet and $\f{\mu}{}$ to the 
vector nonet (with $p^{\mu} \f{\mu}{} = 0$). One can show \cite{huss91b} that
such a relativistic spinor is equivalent to describing the meson as a
quark-antiquark pair both of which are moving at the same velocity as the meson
and are therefore both on-shell.

We use the simplest effective interaction Lagrangian
\beq
 \Li = G \; \Phi_{A}^{B}(\po) \; [\Phi_{B}^{C}(\ptw) \; \Phi_{C}^{A}(\pth)
		+ (\ptw \leftrightarrow \pth) ]	\label{eq:Lint1}
\eeq
as proposed by \cite{delb65a} to describe the three point coupling between the
mesons involved in two-body decays.  $G$ is a normalization factor and we
note that it has the dimensions of mass.  Such an interaction Lagrangian
corresponds to a duality diagram \cite{hara69,rosn69} as shown in Figure~1 with three mesons meeting
at a vertex ($p_i$ incoming).  Here flavour labels have been included to show how the flavour is
automatically conserved, flavour being carried by the line. Contravariant
spinor indices correspond to odd parity as they represent the antiquark, while
covariant indices have even parity since they represent the quark flavour.

Upon substitution of the supermultiplet field (\ref{eq:superfield}) into the
interaction Lagrangian (\ref{eq:Lint1}), and using the Dirac trace algebra
along with the momentum conditions 
\[ \po +\ptw +\pth =0, \;\;\; p_{i}^{2} = m_{i}^{2}, \;\;\; \f{}{}(i) = \f{}{}(p_{i})	\]
we reduce the interaction Lagrangian to 
\bea
 \Li &=& \gvpp \, (p_2 - p_3)^{\mu} < \fm{}{}{1} 
	[\ff{}{}{2}, \ff{}{}{3}] >  \label{eq:Lint21} \\	
     && \!\! + \, \gvvp \,\epsilon^{\mu \nu \kappa \lambda}\,
 	p_{1\kappa}\, p_{2\lambda} \, < \fm{}{}{1}
	\{\fn{}{}{2}, \ff{}{}{3} \} > \label{eq:Lint22} \\ 
     && \!\! + \, \gvvv [(\ptw-\pth)^{\mu} g^{\nu \sigma} \mo + (\pth-\po)^{\nu}
	g^{\sigma \mu} \mtw + (\po-\ptw)^{\sigma} g^{\mu \nu} \mth \nonumber \\
     && \;\;\; + 2 (\ptw-\pth)^{\mu} (\pth-\po)^{\nu} (\po-\ptw)^{\sigma} /
		(\mo+\mtw+\mth)] \nonumber \\
     &&	\;\;\;\;\; < \fm{}{}{1} [\fn{}{}{2}, \phi_{\sigma}(3)] >
	 \label{eq:Lint23}
\eea
where $<>$ stands for a trace over the internal symmetry indices, corresponding
to a joining of quark lines in a duality diagram.  For instance,
such a trace for flavour indices would expand as
\[ \fm{a}{b}{1}	( \ff{b}{c}{2} \ff{c}{a}{3} - \ff{b}{c}{3} \ff{c}{a}{2} ) \]
for the vector--pseudoscalar--pseudoscalar (VPP) vertex. The interaction
Lagrangian contains three distinct coupling constants as at this stage we are
 alert to the possiblility that symmetry breaking may affect each piece 
differently, that is we have introduced three separate constants depending 
on the type of decay.  However, full supermultiplet symmetry would mean the constants are
related in the following way: 
\beq
	2 \gvpp = \mo \gvvp =  \mo \gvvv
\label{eq:predict}
\eeq	
One should also note that the three pseudoscalar meson vertex is not present in
the interaction Lagrangian, as expected by parity conservation. (This follows
automatically in the supermultiplet scheme because the trace over three $\f{5}{}$ fields disappears from 
Equation~\ref{eq:Lint1}.)
	
For the moment we only consider strong meson decay so that tree-level
calculations in perturbation theory given by (\ref{eq:Lint21}, 
\ref{eq:Lint22}, \ref{eq:Lint23}) suffice. Also we apply the general formula
for a two-body decay, 
\beq
 \decay{1}{2,3} = \frac{\lam{1/2}}{16 \pi \mo^{3} (2 s_1 + 1)} 
	\sum_{\rm{spins}}^{} \mod{\Li}^{2}
\label{eq:dec} 
\eeq
where $s_1$ is the spin of the parent meson and $\lam{} = \mo^{4} + \mtw^{4} + \mth^{4} - 2 \mo^{2} \mtw^{2}
 - 2 \mo^{2} \mth^{2} - 2 \mtw^{2} \mth^{2} $.

Upon substitution of the interaction Lagrangian in the form of Equations
(\ref{eq:Lint21},\ref{eq:Lint22},\ref{eq:Lint23}) into the decay rate formula
(\ref{eq:dec}), we derive the following widths for the various decays: 
\bea
 \decay{V}{PP} &=& \lam{3/2} \gvpp^2 / 48 \pi \mo^5	\label{eq:d vpp} \\
 \decay{V}{VP} &=& \lam{3/2} \gvvp^2 / 96 \pi \mo^3	\label{eq:d vvp} \\
 \decay{V}{VV} &=& \lam{3/2} \gvvv^2 {\cal Y}(\mo,\mtw,\mth)/ 192 \pi \mo^5 
                                       \mtw^2 \mth^2,	\label{eq:d vvv}
\eea
where 
\beas
 {\cal Y}(\mo,\mtw,\mth) &=& 9 \left( \sum_{1 \leq i<j \leq 3} (m_i + m_j)^2 (m_i - 
m_j)^4 - \sum_{i=1}^{3} m_i^6 \right) + \\
&& \;\;\; \prod_{i=1}^{3} m_i \left(98 \sum_{i=1}^{3} m_i^3 - 16 \sum_{1 \leq i<j \leq 
3} (m_i + m_j)^3 \right) + 142 \prod_{i=1}^{3} m_i^2.
\eeas

We now have an adequate formalism for describing various strong interaction
 decays amongst ground state mesons.  To extend the applications of the 
supermultiplet theory we invoke the ideas of the vector meson dominance 
model in order to account for various electromagnetic interactions 
of our mesons.  To do so we make the usual assumption that the coupling 
between a vector meson flavour singlet and a photon is of the form 
\beq
	g_{_{V}\g}(k^2 = 0) = e m_{V}^{2} / \gvpp ,
\label{eq:vmd}
\eeq
as shown in Figure~2.  When extrapolating away from 
$k^2 = 0$ we expect the coupling to decrease and as such have denoted the
coupling by $e m_{V}^2 / \gvppd$ to allow for such change.  That is, we
anticipate $\gvppd(k^{2})$ will vary with $k^2$ due to intermediate virtual
particle contributions and its value at $k=0$ equals $\gvvp$ in Equation 
\ref{eq:Lint21}.

The vector meson dominance model, used in conjunction with our decay rate
formulae~(\ref{eq:d vpp},\ref{eq:d vvp},\ref{eq:d vvv}) give the
following rates for the various processes:
\bea
 \decay{V}{l \bar{l}} &=& (m_{V}^2 - 4 m_{l}^2)^{1/2} (1 - m_{l}^2 / m_{V}^2)^{1/2}
	(e^2/\gvppd)^2 /12 \pi	\label{eq:d vll}	\\
 \decay{V}{P\g} &=& (m_{V}^2 - m_{P}^2)^3 (e \gvvp / \gvppd)^2 / 96\pi m_{V}^3
	\label{eq:d vpg}	\\
 \decay{P}{V\g} &=& (m_{P}^2 - m_{V}^2)^3 (e \gvvp / \gvppd)^2 / 32\pi m_{P}^3
	\label{eq:d pvg}	\\
 \decay{P}{\g\g} &=& m_{P}^3 (e^2 \gvvp / \gvppd \gvppd)^2 / 64 \pi
	\label{eq:d pgg}	
\eea
thereby greatly extending the original scope of the supermultiplet scheme.  
In going from our purely strong interaction decay rates to the radiative ones, we
have used the gauge invariance of our interaction Lagrangian and simply
substituted a mass of zero for those vectors connecting with the photon. 
However, the three vector interaction (\ref{eq:Lint23}) is only gauge 
invariant for the case $\mtw = \mth$ so strictly we should only apply it 
to radiative examples for which the virtual vector meson satisfies this 
condition.  Unfortunately, since the photon only couples to flavour singlet  
states the condition $\mtw = \mth$ also implies the daughter vector mesons are 
identical.  Due to the F--type coupling between daughter states in the interaction 
Lagrangian (\ref{eq:Lint23}) such decay widths will automatically go to zero.  
It is for this reason we have not included a $V \to V \g$ term above, despite 
experimental evidence for such (eg. $\decay{\phi}{\rho \g}/\decay{\phi}{\rm{all}}
 < 2\%$; although our zero width prediction does not conflict with this). 
We now go one to apply the formalism to the ground state mesons.

\section{Supermultiplet Method}

In the standard way we take the pseudoscalar nonet as: 
\beq
	\phi_{5 a}^{\;\; b} \stackrel{0^-}{\rightarrow}
\left( \begin{array}{ccc}
\tfrac{\pi^0}{\sqrt{2}}+\tfrac{\eta_8}{\sqrt{6}}+\tfrac{\eta_{0}}{\sqrt{3}} &
	\pi^+	&	K^+	\\
	\pi^- & -\tfrac{\pi^0}{\sqrt{2}}+\frac{\eta_8}{\sqrt{6}}
+\frac{\eta_0}{\sqrt{3}} & K^0	\\
 K^- & \bar{K}^{0} &  -\frac{2\eta_8}{\sqrt{6}}+ \frac{\eta_0}{\sqrt{3}}
 \end{array} \right)	\label{matrix:pseudo}
\eeq
where 
\bea
 \eta_8 &=& \cos \tP \;\eta - \sin \tP \;\eta' 	\label{eq:e81} \\
 \eta_0 &=& \sin \tP \;\eta + \cos \tP \;\eta'	\label{eq:e01}
\eea
as defined in \cite{gasi} and $\tP$ is the pseudoscalar mixing angle.
The vector nonet is similarly given by
\beq
	\phi_{\mu a}^{\;\; b} \stackrel{1^-}{\rightarrow}
\left( \begin{array}{ccc}
\frac{\rho^{0}}{\sqrt{2}}+\frac{\omega_{8}}{\sqrt{6}}+\frac{\omega_{0}}{\sqrt{3}} &
	\rho^{+}	&	K^{*+}	\\
	\rho^{-} & -\frac{\rho^{0}}{\sqrt{2}}+\frac{\omega_{8}}{\sqrt{6}}
+\frac{\omega_{0}}{\sqrt{3}} & K^{*0}	\\
 K^{*-} & \bar{K}^{*0} &  -\frac{2\omega_{8}}{\sqrt{6}}+ \frac{\omega_{0}}{\sqrt{3}}
 \end{array} \right)   \label{matrix:vector}
\eeq
where
\bea
 \omega_{8} &=& \cos \tV \;\phi - \sin \tV \;\omega	\label{eq:w81} \\
 \omega_{0} &=& \sin \tV \;\phi + \cos \tV \;\omega \label{eq:w01}
\eea
and $\tV$ is the vector mixing angle as determined using the Gell-Mann--Okubo
(GMO) mass relation.  We have chosen to go into some detail as
our formalism is not the common one adopted by recent publications
\cite{bram9001,pdg}. Conversely,
\beq
  \phi = \cos \tV \:\omega_{8} + \sin \tV \:\omega_{0} \label{eq:phi}
\eeq
and the $\omega$ field as 
\beq
  \omega = -\sin \tV \:\omega_{8} + \cos \tV  \:\omega_{0},	\label{eq:omega}
\eeq
where $\omega_8$ and $\omega_0$ masses are determined by the GMO
relation \cite{gasi}. The vector mixing angle is obtained from
\beq
 \tan 2 \tV = \frac{2((m_{\phi}^2 - m_8^2)(m_8^2 - m_{\omega}^2))^{1/2}}{2
m_{8}^{2}- m_{\phi}^{2} - m_{\omega}^2}, \label{eq:theta}
\eeq
where $3 m_{8}^{2} = 4 m_{K^*}^{2} - m_{\rho}^{2}$.  

From the vector nonet (\ref{matrix:vector}) the octet and singlet fields are
expressed in terms of the supermulitplet vector as
\[
 \sqrt{6} \; \omega_{8}= \phi_{\mu 1}^{\;\; 1}+\phi_{\mu 2}^{\;\; 2} -
		2\phi_{\mu 3}^{\;\; 3},	\;\;\;\;\;
 \sqrt{3} \; \omega_{0} = \phi_{\mu 1}^{\;\; 1}+\phi_{\mu 2}^{\;\; 2} +
		\phi_{\mu 3}^{\;\; 3}.	
\]
Substituting these into Equation~\ref{eq:phi} yields
\[
 \sqrt{6} \; \phi= ( \cos \tV + \sqrt{2} \sin \tV) (\phi_{\mu 1}^{\;\; 1}+\phi_{\mu
2}^{\;\; 2})+ (-2\cos \tV + \sqrt{2} \sin \tV) \phi_{\mu 3}^{\;\; 3}.
\]
In the case of ``ideal mixing'' $\phi = \phi_{\mu 3}^{\;\; 3}$ so that 
\[	\cos \tV + \sqrt{2} \sin \tV = 0 \;\;\;\; \mbox{or} \;\;\;\;
	\tan \tV = -1/\sqrt{2}	\]
leaving us two options for $\tV$; either $-\pi/2 < \tV < 0$ or $\pi/2< \tV < \pi$.
The first case implies $\cos \tV=\sqrt{2/3}, \;\; \sin \tV =- 1/\sqrt{3}$ so that
$\phi = -\phi_{\mu 3}^{\;\; 3}$ while the second gives the desired result of
$\phi = \phi_{\mu 3}^{\;\; 3}$.  Thus a suitable solution to Equation
\ref{eq:theta} is in the range $\pi/2 < \tV < \pi$.  More generally, the
solution to (\ref{eq:theta}) is 
\[ 2 \tV = \tan^{-1} \left( \frac{2((m_{\phi}^2 - m_8^2)(m_8^2 - m_{\omega}^2))^{1/2}}{2
m_{8}^{2}- m_{\phi}^{2} - m_{\omega}^2} \right) \; + \; n \pi	\] 
where $n$ is any integer.

The above arguments for the determination of the vector mixing angle can be
applied to the pseudoscalar nonet with the substitutions $\omega_8 \to \eta_8,
\; \omega_0 \to \eta_0, \; \phi \to \eta,
\; \omega \to \eta'$.  Using the condition $\pi/2 < \tV < \pi$ and a similarly
derived expression for the pseudoscalar angle, $-\pi/2 < \tP < \pi/2$, we
obtain the equally likely results 
\beas
	\tV &=& \dg{129.4}, \; \dg{140.6} 	\\
	\tP &=& \dg{-10.5}, \; \dg{10.5}.
\eeas
The mixing angles we have obtained may seem accurate, but the GMO
relation is extremely sensitive to an extra small SU(3) symmetry breaking
associated with the $\underline{27}$ representation; a small 
$\underline{27}$ addition can produce a {\em major} modification of the angle.

With the correct structure now in place, it is a relatively simple process to
test the supermultiplet theory.  We wish to calculate the standard coupling constants
$\gvpp$ and $\gvvp$,  examine how similar they are for each process and
finally compare the supermultiplet prediction (\ref{eq:predict}) of the 
relation between them.  In
practice we take the decay width and particle masses as input \cite{pdg} and determine the coupling
constant associated with the decay via (\ref{eq:d vpp},\ref{eq:d vvp}) and 
(\ref{eq:d vll}--\ref{eq:d pgg}).  The simplicity of the supermultiplet method is
that isospin and mixing factors are automatically accounted for.  One simply chooses
an appropriate decay, determines the flavour indices
$a,b$ and $c$ using matrices (\ref{matrix:pseudo},\ref{matrix:vector}) and then
use these in the correct part of the interaction Lagrangian (\ref{eq:Lint21},
\ref{eq:Lint22}, or \ref{eq:Lint23}) to determine the normalization factors
which arise.  For example, in the decay $\rho^+ \to \pi^+ \: \pi^0, \; a=1, 
\:b=2, \: c=1,2$ and upon substitution of the fields into
Equation~\ref{eq:Lint21} one finds $g_{\rho^+ \pi^+ \pi^0} = \sqrt{2} \; \gvpp$
so that the coupling constant we determine for this decay should be divided by
the factor $\sqrt{2}$ to obtain the standard coupling constant $\gvvp$.  This
procedure is repeated for all appropriate physical decays.  Mixing is easily
accommodated by using the relations
(\ref{eq:e81},\ref{eq:e01},\ref{eq:w81},\ref{eq:w01}) to replace the ideal
fields by the real mesons in the interaction Lagrangian.   

In radiative decays of the type $V \to l \; \bar{l}$ we allow for the coupling
of the photon to the quark. Using the following electromagnetic charge
projectors 
\beq
 Q_{a}^{b} = \left( \begin{array}{ccc}
		2/3 	&	0	&	0	\\
		0	&	-1/3	&	0	\\
		0	&	0	&	-1/3
	\end{array}	\right)
\eeq
we may likewise extract the relevant standard coupling.  

Radiative modes such as $V \to P \g$ require some delicacy in normalising
the coupling constant.  To elicit a clear understanding of the method
we include an example of the procedure for the decay $\rho^0 \to \eta \g$.  
Firstly, we recognise $\rho^0$ is a combination of $\fm{1}{1}{1}$ and 
$\fm{2}{2}{1}$, so we require terms in the interaction Lagrangian 
(\ref{eq:Lint22}) with $a=1, \: b=1$ and $a=2, \: b=2$.  For each of these 
cases we determine the third flavour index $c$ such that $\fn{}{}{2}$ is a 
flavour singlet that couple to a photon (thus $c=1, 2$).  Substituting these 
values into formula \ref{eq:Lint22} we pick out the uncharged parts;
\beas
 \Li & \propto & \gvvp \; [ \; \fm{1}{1}{1} \; \{ \fn{1}{1}{2},\ff{1}{1}{3} \} 
	 + \fm{2}{2}{1} \; \{ \fn{2}{2}{2},\ff{2}{2}{3}\} \;] \\
	&=& \gvvp \; [ \tfrac{\rho^0 (1)}{\sqrt{2}} (2 (\tfrac{\rho^0 (2)}{\sqrt{2}} +
\tfrac{\omega_{8}(2)}{\sqrt{6}} ) (\tfrac{\eta_{8}(3)}{\sqrt{6}} +
\tfrac{\eta_{0}(3)}{\sqrt{3}}) \\
	& & \;\;\; \tfrac{-\rho^0 (1)}{\sqrt{2}} (2 (\tfrac{-\rho^0 (2)}{\sqrt{2}} +
\tfrac{\omega_{8}(2)}{\sqrt{6}} ) (\tfrac{\eta_{8}(3)}{\sqrt{6}} +
\tfrac{\eta_{0}(3)}{\sqrt{3}})] \\
	&=& \sqrt{\tfrac{2}{3}} \: \gvvp \: \rho^0 (1) \: \rho^0 (2) \: [\eta_8 (3) +
\sqrt{2} \eta_0 (3)] \\
	&=& \sqrt{\tfrac{2}{3}} (\cos \tP + \sqrt{2} \sin \tP) \; \gvvp \; \rho^0 (1)
\; \rho^0 (2) \; \eta (3),
\eeas 
where in particular we have used relations (\ref{eq:e81},\ref{eq:e01}) to
arrive at the final result.  The form shows that the coupling between two 
$\rho^0$ mesons and a pseudoscalar $\eta$ is related to the standard VVP 
coupling by 
\beq
  g_{\rho^0 \rho^0 \eta} = \sqrt{\tfrac{2}{3}} (\cos \tP + \sqrt{2} \sin \tP) \; \gvvp 
\label{eq:grre}
\eeq 
The virtual vector meson is immediately identifiable as $\rho^{0}(2)$, and 
we must necessarily allow for the coupling between this and the photon.
Since $\rho^0 = (u \bar{u} + d \bar{d})/\sqrt{2}$ then $g_{\rho^0 \g} =
g_{_{V}\g}/\sqrt{2}$ which in turn implies $g_{\rho^0 PP}' = \sqrt{2}
\gvppd$ from (\ref{eq:vmd}).
Subsequently, the coupling between a $\rho^{0}, \: \eta$ and photon is
related to our standard couplings by
\beas
  g_{\rho^0 \eta \g} &=& e  g_{\rho^0 \rho^0 \eta} / g_{\rho^0 PP}'
\\
  	&=& \frac{e \: \sqrt{\tfrac{2}{3}} (\cos \tP + \sqrt{2} \sin \tP) \gvvp}
{\sqrt{2} \gvppd} \\
	&=& \frac{1}{\sqrt{3}} \: (\cos \tP + \sqrt{2} \sin \tP) \; 
	\frac{e \gvvp}{\gvppd}
\eeas

In other decays, it is possible that the radiative mode may proceed via more 
than one virtual vector meson.  The above method is still used to determine 
each virtual vector meson contribution and the appropriate linear combination
is taken.

\section{Results}

The results are presented in Tables 1 and 2.  For clarity these tables 
include the SU(3) factors which we have used to normalize the coupling constant.  

Table 1 summarises the results of our investigation into the coupling between a
vector meson and two pseudoscalar mesons.  The first half of Table 1 displays
purely strong interaction decays, while the second lists the coupling
constant $\gvppd$ obtained from vector meson dominance extrapolation. 
Two important features are apparent.  Firstly, we have found
that the coupling is far more regular than previously believed by those persons
who deprecate light quark symmetry.  Secondly, the
form of the symmetry breaking is now very clear.  As the mass of the parent
vector meson increases (as we go down each half of the table), so does the
coupling, apparently following the simple rule $\gvpp \approx 0.154 \: m_V^{1/2}$ 
 (for $m_V$ in units MeV).  Similarly,
the mass-shell constants $\gvppd$ follow such a relation, except the constant
of proportionality is approximately $(0.136 \pm 0.003) \rm{MeV}^{-1/2}$ by a weighted mean method 
(and an error scale factor of 4; following the Particle Data Group's handling of
 errors). This result complies with the known 
scaling law behaviour for $f_V$ as $m_V \to \infty$ \cite{grin92}.  Thus if the 
matrix elements of quark currents between a given vector meson and the vacuum 
state is defined by
\[	<0 \mod{\bar{q}_2 \gm q_1} V > = f_V m_V \epsilon_{\mu}
\]
then it is well known \cite{rein} that $f_V \propto \mod{\psi(0)} /
m_V^{1/2}$ as $m_V \to \infty$.  Translating to our terminology 
$f_V = e m_V / \gvppd$, we verify this prediction and importantly we find 
the result is also supported in the light meson sector.  Admittedly, the 
$\omega \to e^+ e^-$ has a very high $\gvppd$, but since the width of $\omega
\to \mu^+ \mu^-$ is only known to an upper bound (providing a lower bound
estimate of $\gvppd$) the anomaly remains unsubstantiated.

Table 2 predominantly lists the results from studying radiative decays to
obtain estimates of $\gvvp$ using vector meson dominance. The first entry in
the table is for the decay $\phi \to \rho \pi$ and leads to a direct
determination $\gvvp$, not via a radiative transition.  In fact, we use it to
test the supermultiplet prediction $2 \,\gvpp = \mo \,\gvvp$, the results of which
are shown in Figure~3.  This figure shows the sum $\mo \gvvp - 2 \gvpp$ plotted
against vector mixing angle $\tV$ and clearly demonstrates the supermultilpet
rule is satisfied at $\tV \approx \dg{140.3}$, very close to the accepted value
$\tV = \dg{140.6}$, and it is for this reason we have used this value in all
our calculations.   

In the case of radiative decays, where we know the coupling is related to the
ratio $\gvvp/\gvppd$, we have used the relationship $\gvppd \approx 0.136 \:
m_V^{1/2}$, which is well supported by the data in Table~1.  Importantly, this
relation applies to the virtual vector meson so that for decays mediated via
the ideal field $\omega_8$ we have to use its mass of approximately 931 MeV.
The data shows that once again the coupling is quite regular, but now the
symmetry breaking appears to obey a power law relation $\gvvp \propto 
m_1^{-n}$ where $1/2 < n < 3/2$.

\section{Predictions}
 
With a clearer understanding of the effects of symmetry breaking on the 
coupling constant, we may now confidently determine the decay rates for 
non-Zweig allowed decays.  In particular we study the decays $\omega \to 
\pi^+ \pi^-$ and $\phi \to \pi^+ \pi^-$, both of which are mediated by a virtual 
photon coupling between the parent vector meson and a $\rho$ meson
(electromagnetic mixing). Thus in the case $\omega \to \pi^+ \pi^-$ we have the
overall coupling of 
\[
 g_{\omega\pi\pi} = e^2 m_{\rho}^2 g_{\rho\pi\pi} / g_{\omega _{PP}}' 
	(m_{\omega}^2 - m_{\rho}^2) g_{\rho _{PP}}'     
\]
and for $\phi \to \pi^+ \pi^-$ we have 
\[ 
g_{\phi\pi\pi} = e^2 m_{\rho}^2 g_{\rho\pi\pi} /  g_{\phi PP}' 
	(m_{\phi}^2 - m_{\rho}^2) g_{\rho _{PP}}'.      
\]
Using 
\beas
   g_{\rho\pi\pi} &=& \sqrt{2}\: (0.1537 \pm 0.002) m_{\rho}^{1/2}  \\
   g_{\rho _{PP}}' & = & \sqrt{2}\: (0.136 \pm 0.003) m_{\rho}^{1/2}  \\
   g_{\omega _{PP}}'&=& \sqrt{6}\: (0.136 \pm 0.003) m_{\omega}^{1/2} / \sin \tV \\
   g_{\phi _{PP}}' &=& \sqrt{6}\: (0.136 \pm 0.003) m_{\phi}^{1/2} / \cos \tV 
\eeas
we predict
\beas
   \decay{\omega}{\pi^+ \pi^-} &=& (1.66 \pm 0.16) \times 10^{-2}  \; \rm{MeV}  \\
   \decay{\phi}{\pi^+ \pi^-} &=& (5.88 \pm 0.55) \times 10^{-4} \; \rm{MeV}
\eeas
which compare favourably with the presently accepted values
\beas
   \decay{\omega}{\pi^+ \pi^-} &=& (1.86 \pm 0.25) \times 10^{-2}  \; \rm{MeV}  \\
   \decay{\phi}{\pi^+ \pi^-} &=& (3.5 \pm 2.8) \times 10^{-4} \; \rm{MeV}.
\eeas

The symmetry breaking effects we have observed also lead to a 
measurable consequence in the radiative decays of heavy mesons.  We begin by
re-examining the decays $K^{*\pm} \to K^{\pm} \g$ and $K^{*0} 
\to K^0 \g$.  Experimentally, the $K^*$ branching fraction is
\[ 
\decay{K^{*0}}{K^0 \g} / \decay{K^{*+}}{K^+ \g} = 2.31 \pm 0.29 
\]
and allowing for phase space factors this translates into a coupling
constant ratio of
\[ 
 \mod{ g_{K^{*0} K^0 \g}/g_{K^{*+} K^+ \g}}  = 1.514 \pm 0.095,	\]
and as such is far from the exact SU(3) ratio of 2.   Under the
supermultiplet scheme, one can show the decays proceed via two
intermediate vector mesons, $\rho^0$ and $\omega_8$.   Following the
procedure we described for determining the normalisation factors, one finds
\bea
 g_{K^{*+} K^+ \g} &=& e \left( \frac{g_{K^{*+} \rho^0 K^+}}{g_{\rho^0
	PP}'} + \frac{g_{K^{*+} \omega_8 K^+}}{g_{\omega_8 PP}'} \right)
\label{eq:K+g} \\
 g_{K^{*0} K^0 \g} &=& e \left( \frac{g_{K^{*0} \rho^0 K^0}}{g_{\rho^0
	PP}'} + \frac{g_{K^{*+} \omega_8 K^0}}{g_{\omega_8 PP}'} \right).
\label{eq:K0g}
\eea
If one assumes $\gvppd$ is constant then
\beas
 g_{K^{*+} K^+ \g} &=& e \left( \frac{1/\sqrt{2}}{\sqrt{2}} + 
	\frac{-1/\sqrt{6}}{\sqrt{6}}  \right) (\gvvp/\gvppd) \\
	&=& \gvpg / 3 \\
 g_{K^{*0} K^0 \g} &=& e \left( \frac{-1/\sqrt{2}}{\sqrt{2}} + 
	\frac{-1/\sqrt{6}}{\sqrt{6}}  \right) (\gvvp/\gvppd) \\
	&=& - 2 \gvpg / 3 ,
\eeas
and we arrive at the exact SU(3) prediction.  If instead we use a symmetry
 breaking $\gvppd$ we must
substitute $g_{\rho^0 PP}' = \sqrt{2} \, C \, m_{\rho^0}^{1/2}$ and 
$g_{\omega_{8} PP}' = \sqrt{6} \, C \, m_{\omega_8}^{1/2}$ in Equations
(\ref{eq:K+g}) and (\ref{eq:K0g}).  Thus
\[ 
\frac{g_{K^{*0} K^0 \g}}{ g_{K^{*+} K^+ \g}} = - \, \frac{m_{\rho^0}^{-1/2} +
m_{\omega_8}^{-1/2}/3 }{m_{\rho^0}^{-1/2} - m_{\omega_8}^{-1/2}/3 } = -
1.87
\]
and notice the result is independent of $C$, the constant of 
proportionality between $\gvppd$ and $m_V^{1/2}$.  Although not matching 
the experimental result, it is an improvement on exact SU(3). Actually, the most 
satisfactory explanation of the symmetry breaking mechanism comes from 
\cite{bram8912}. They attribute the deviation from exact SU(3) to the 
constituent mass difference between the strange and non-strange quarks
in the loop of a quark triangle diagram.  As $K^* \to K \g$ excite both 
strange and non-strange quarks, such a difference must be accounted for.
With these corrections, the experimental ratio is found to match 
theoretical estimates very well.  We intend to apply the method to heavier 
meson cases in future work \cite{liu9409}.
 
Let us continue to use the ``mass variation principle" of $\gvppd$ in the heavy meson 
sector.  Upon application to the $D^*$ and $B^*$ mesons we obtain
\bea
 \frac{g_{D^{*0} D^0 \g}}{ g_{D^{*+} D^+ \g}} &=& \frac{3 m_{\rho^0}^{-1/2} +
m_{\omega_8}^{-1/2} + 4 m_{J/\psi}^{-1/2} }
{- 3 m_{\rho^0}^{-1/2} + m_{\omega_8}^{-1/2} + 4 m_{J/\psi}^{-1/2} } \approx -
60  \label{eq:coupleD} \\
\frac{g_{B^{*0} B^0 \g}}{ g_{B^{*+} B^+ \g}} &=& \frac{-3 m_{\rho^0}^{-1/2} +
m_{\omega_8}^{-1/2} + 2 m_{\upsilon}^{-1/2} }
{3 m_{\rho^0}^{-1/2} + m_{\omega_8}^{-1/2} + 2 m_{\upsilon}^{-1/2} } \approx 
- 0.34   \nonumber
\eea
which are significantly different from the exact SU(5) predictions of
\beas
 g_{D^{*0} D^0 \g}/ g_{D^{*+} D^+ \g} &=& 4 \\
 g_{B^{*0} B^0 \g}/ g_{B^{*+} B^+ \g} &=& 0   
\eeas
and as such require better experimental data to test the results.

In addition to these relative decay rate predictions, the supermultiplet 
scheme can be easily adapted to decay width calculations.  Scattered amongst 
Tables 1 and 2 are various constants determined by extending the supermultiplets to include 
the charm and bottom quark mesons.  In particular, using the upper bound
of 131 KeV for the $D^{*+}$ decay width \cite{accm92} we have
found $\gvpp < 10$.   Conversely, we can use our knowledge of the
effects of symmetry breaking to predict the VPP coupling constant for
$D^{*+}$.  We find
\beq
	\gvpp (D^{*+})\approx (0.1537 \pm 0.002) (2010)^{1/2} = 6.89 \pm 0.09
\label{eq:D*gvpp}
\eeq
surprisingly similar to a heavy quark prediction of $7 \pm 1$ by 
\cite{cola9406}.  
We can use the $\gvpp$ value for $D^{*+}$ to calculate the total decay width 
of the $D^{*+} \to PP$ channels.  Using the supermultiplet method we can 
predict all the possible decays of $D^{*+}$ into two pseudoscalars; however, 
phase space restricts the processes to $D^{*+} \to D^0 \pi^+$ and 
$D^{*+} \to D^+ \pi^0$ so that the width must be
\beas
 \decay{D^{*+}}{PP} &=& \decay{D^{*+}}{D^+ \pi^0} + \decay{D^{*+}}{D^0 \pi^+} \\
	&=& \gvpp^2 [ \lambda^{3/2} (m_{D^{*+}}^{2},m_{D^0}^2,m_{\pi^+}^2) +
      \lambda^{3/2}(m_{D^{*+}}^{2},m_{D^+}^2,m_{\pi^0}^{2})] 
      /48 \pi m_{D^{*+}}^{5}	\\
	&=&  57.7 \pm 1.5 \; \rm{KeV}.
\eeas
We compare this with the radiative width $D^{*+} \to D^+ \g$ in the 
following branching fraction:
\beas
   \frac{\decay{D^{*+}}{D^+ \g}}{\decay{D^{*+}}{PP}} 
 &=& \frac{(\gvpg/\gvpp)^{2} (m_{D^{*+}}^2 - m_{D^+}^2 )^3 /96 \pi 
      m_{D^{*+}}^3  }{[ \lambda^{3/2}
      (m_{D^{*+}}^{2},m_{D^0}^2,m_{\pi^+}^2) +
      \lambda^{3/2}(m_{D^{*+}}^{2},m_{D^+}^2,m_{\pi^0}^{2})]/48 \pi 
      m_{D^{*+}}^5 } \\
 &\approx& \frac{( e (- 3 m_{\rho^0}^{-1/2} + m_{\omega_8}^{-1/2} + 4 
      m_{J/\psi}^{-1/2}) / (6 \times 0.1361))^2 (m_{D^{*+}}^2 - m_{D^+}^2 )^3 
      }{ \lambda^{3/2} (m_{D^{*+}}^{2},m_{D^0}^2,m_{\pi^+}^2) +
      \lambda^{3/2}(m_{D^{*+}}^{2},m_{D^+}^2,m_{\pi^0}^{2}} \\
 &\approx& (9.48 \pm 0.43) \times 10^{-5},
\eeas
where in particular we have used the supermultiplet prediction $\gvvp / 
\gvpp = 2 / m_{D^{*+}}$ and we have confidence in our prediction 
to this order. This finding implies that the dominant decay 
modes in $D^{*+}$ decay are the PP channels we derived and as such 
approximate well to the full width.  Thus the radiative branching 
fraction for $D^{*+} \to D^+ \g$ is relatively small.  This is not inconsistent with 
recent measurements by CLEO which measured a fraction of $(1.1 \pm 1.4 \pm 1.6)\%$ 
\cite{cleo92a}.  However, our prediction does conflict with other 
theoretical models \cite{kama92,cola9407,odon9407,jain9407}. Branching 
fraction calculations for the two PP channels yield 
\beas
 \decay{D^{*+}}{D^0 \pi^+}/ \decay{D^{*+}}{\rm{all}} &\approx& 68.8 \%  \\
 \decay{D^{*+}}{D^+ \pi^0}/ \decay{D^{*+}}{\rm{all}} &\approx& 31.2 \%  
\eeas
which compare well with other models and the experimentally determined results 
from CLEO (1992):
\beas
 \decay{D^{*+}}{D^0 \pi^+}/ \decay{D^{*+}}{\rm{all}} &=& (68.0 \pm 1.4 \pm 2.4)\%  \\
 \decay{D^{*+}}{D^+ \pi^0}/ \decay{D^{*+}}{\rm{all}} &=& (31.0 \pm 0.4 \pm 1.6)\%.  
\eeas

We are able to employ similar methods in the decays of the $D^{*0}$
vector meson.  In this instance, the possible PP decay channels are restricted 
by phase space to $D^{*0} \to D^0 \pi^0$.  We calculate this width to be
$27.2 \pm 0.7$ KeV, where we used $g_{D^{*0} D^0 \pi^0} = (0.1537 \pm
0.002) (2007.1)^{1/2} / \sqrt{2}$.  To determine the radiative width
$D^{*0} \to D^0 \g$ we apply relation (\ref{eq:coupleD}) along with a
small correction for the change in phase space to derive
\[
 \decay{D^{*0}}{D^0 \g} \approx 3672 \times \decay{D^{*+}}{D^+ \g} 
	= (20.1 \pm 1.1) \;\; \rm{KeV}.
\]
Thus the total $D^{*0}$ width is $(47.3 \pm 1.3)$ KeV, although the result is
sensitive to the supermultiplet prediction $\gvvp(D^*)/\gvpp(D^*) =
2/m_{D^*}$.  Consequently,  we predict the following branching fractions
\beas
 \decay{D^{*0}}{D^0 \pi^0}/ \decay{D^{*0}}{\rm{all}} &\approx& (57.5 \pm 2.2) \%  \\
 \decay{D^{*0}}{D^0 \g}/ \decay{D^{*0}}{\rm{all}} &\approx& (42.5 \pm 2.6) \%  
\eeas
which are in fair agreement with the CLEO data:
\beas
 \decay{D^{*0}}{D^0 \pi^0}/ \decay{D^{*0}}{\rm{all}} &=& (64 \pm 2.4 \pm 4.5) \%  \\
 \decay{D^{*0}}{D^0 \g}/ \decay{D^{*0}}{\rm{all}} &\approx& (36 \pm 2.4
	\pm 4.5) \%.
\eeas

\section{Conclusions}
 
This study of two-body meson decays has shown that the supermultiplet 
method unifies meson decays quite well, even for the light quarks.  
The most significant finding is that the coupling between mesons is 
susceptible to symmetry breaking mechanisms, but in a {\em regular} way, 
allowing us to successfully extrapolate to decay rates for other 
processes.  In particular, the methods are readily applicable to heavy 
quark examples, as highlighted by our examination of $D^*$ processes.

\vspace{.2in}
\noindent
{\bf Acknowledgements}

The authors would like to thank Dr.\ Dongsheng Liu for many discussions along
with Dr.\ Dirk Kriemer for insights into extrapolating the vector dominance
model from $k^2 = 0$ to other values.


\newpage
\pagestyle{empty}

\noindent
{\bf \Large Table captions.}
\begin{description}
\item[Table 1] Results for $\gvpp$ determination.
\item[Table 2] Results for $\gvvp$ determination.

\end{description}
\vspace*{4cm}

\noindent
{\bf \Large Figure captions.}
\begin{description}
\item[Figure 1] Quark line or duality diagram.
\item[Figure 2] Vector meson dominance.
\item[Figure 3] Support for the supermultiplet symmetry condition $2 \gvpp =
	\mo \gvvp$ in the decays $\phi \to K^0 K^0$ and $\phi \to \rho \pi$.

\end{description}

\newpage
\begin{center}
\begin{tabular}{|l|c|l|}
\hline \hline
 Decay			&Factor & \large $\gvpp$ \normalsize \\
\hline
 $\rho^{\pm} \to \pi^{\pm}\pi^0$&$\sqrt{2}$& $4.24 \pm 0.05$ \\
 $\rho^0 \to \pi^+\pi^- $	&$\sqrt{2}$& $4.30 \pm 0.03$   \\
 $\rho^{\pm} \to \pi^{\pm}\eta$	&$\sqrt{1/6}(\cos \tP+\sqrt{2}\sin \tP)$	& $< 4.17\pm 0.16$		   	\\
\hline
 $K^{*\pm} \to (K \pi)^{\pm}$	&$ 1\, ,\;\; \sqrt{1/2}$& $4.59$	   	\\
 $K^{*0} \to (K \pi)^0$	&$\sqrt{1/2}\, ,\;\; 1 $& 4.55		   	\\
\hline
 $\phi \to K^+K^-$	&$\sqrt{3/2}\cos \tV$& $4.82 \pm 0.05$  	\\
 $\phi \to K^{0}_{L}K^{0}_{S}$&$\sqrt{3/2}\cos \tV $& $4.99 \pm 0.06$  	\\
\hline
 $D^{*+} \to D^0 \pi^+$	& 1	& $<10.2 \pm 1.0$  	\\
 $D^{*+} \to D^+ \pi^0$	& $1/\sqrt{2}$	& $<10.3 \pm 1.1$  \\
\hline \hline
 Decay	& Coupling factor & \large $\gvppd$ \normalsize  \\
\hline
$\rho^0 \to e^+ e^-$	&$\sqrt{1/2}$	&$3.57\pm0.09$ \\
 $\rho^0 \to \mu^+ \mu^-$&$\sqrt{1/2}$	&$3.41\pm0.11$ \\
\hline
 $\omega \to e^+ e^-$	&$\sqrt{1/6}\sin \tV$&$4.40\pm0.06$ \\
 $\omega \to \mu^+ \mu^-$&$\sqrt{1/6}\sin \tV$&$>2.70$ \\
\hline
 $\phi \to e^+ e^-$	&$\sqrt{1/6}\cos \tV$&$4.07\pm0.05$ \\
 $\phi \to \mu^+ \mu^-$	&$\sqrt{1/6}\cos \tV$&$4.46\pm0.31$ \\
\hline
 $J/\psi \to e^+ e^-$	&2/3&	$7.55\pm0.29$ \\
 $J/\psi \to \mu^+ \mu^-$&2/3&	$7.72\pm0.31$ \\
\hline
 $\Upsilon \to e^+ e^-$	&1/3&	$13.36\pm0.53$ \\
 $\Upsilon \to \mu^+ \mu^-$	&1/3&	$13.47\pm0.32$ \\
 $\Upsilon \to \tau^+ \tau^-$ &1/3& $11.63\pm0.72$ \\
\hline\hline
\end{tabular}
\hspace*{8.1cm} \tiny{$\tV = \dg{140.6}, \; \tP = \dg{10.5}$}
\normalsize \\
Table 1.
\end{center}

\begin{center}
\begin{tabular}{|l|c|l|}	\hline \hline
 Decay			&Factor & \large $g_{_{VVP}}$ \normalsize \\ 
               &       & $\times 10^{-2}$ MeV$^{-1}$  \\\hline
 $\phi\to\rho\pi$&\sm{$\sqrt{2/3}(\cos \tV+\sqrt{2}\sin \tV)$}&$1.062\pm0.030$  \\
\hline \hline
 $\rho^{\pm}\to\pi^{\pm}\g$ & $1/3$ & $0.923\pm0.055$ \\
 $\rho^0 \to \pi^0 \g$ & $1/3$ & $ 1.216 \pm 0.156$  \\
 $\rho^0 \to \eta \g$  & $(\cos \tP+\sqrt{2}\sin \tP)/\sqrt{3}$ & $0.984\pm0.094$ \\
\hline
 $\omega \to \pi^0 \g$ & $(\cos \tV - \sin \tV)/\sqrt{3}$ & $0.878\pm0.033$ \\
 $\omega \to \eta \g$  & $(\sqrt{2}\cos (\tV+\tP) + \sin \tV \cos \tP)/3$ & $0.919\pm0.165$ \\     
\hline
 $\phi \to \pi^0 \g$   & $(\cos \tV + \sqrt{2}\sin \tV)/\sqrt{3}$ & $0.731\pm0.040$ \\
 $\phi \to \eta \g$    & $(\sqrt{2}\sin (\tV+\tP) - \cos \tV \cos \tP)/3$ & $0.603\pm0.020$ \\
 $\phi \to \eta'\g$   & $(\sqrt{2}\cos (\tV+\tP) +  \cos \tV \sin \tP)/3$ & $<1.69$ \\     
\hline
 $K^{*\pm}\to K^{\pm}\g$& $1/3$ & $0.905 \pm 0.045$ \\
 $K^{*0} \to K^0 \g$   & $-2/3$ & $0.733 \pm 0.036$ \\
\hline
 $J/\psi\to\eta_c \g$ & $4/3$ & $0.308\pm0.073$   \\
\hline \hline
 $\eta' \to \rho^0 \g$ & $(\sqrt{2} \cos \tP - \sin \tP)/\sqrt{3}$ & $0.693\pm0.041 $ \\
 $\eta' \to \omega \g$ & $-(\sin \tP \sin \tV + \sqrt{2}\sin (\tP+\tV))/3$ & $0.698\pm0.051$  \\
\hline \hline
 $\pi^0 \to \g\g$ & $ \sqrt{2}/3$ & $ 0.915\pm0.053$  \\
 $\eta \to \g\g$  & $ \sqrt{2/3} (\sqrt{2}\sin \tP + \cos \tP)/3$ & $0.870\pm0.056$ \\
 $\eta' \to \g\g$ & $ \sqrt{2/3} (\sqrt{2}\cos \tP -\sin \tP)/3$ & $0.730\pm0.055$ \\
 $\eta_c \to \g\g$& $ 8/9 $ & $0.484\pm0.337$   \\
\hline
\hline
\end{tabular}
\hspace*{9.3cm} \tiny{$\tV = \dg{140.6}, \; \tP = \dg{10.5}$}
\normalsize \\
Table 2.
\end{center}

\newpage 
\hspace*{1cm} \epsfxsize = 13cm	\epsfbox{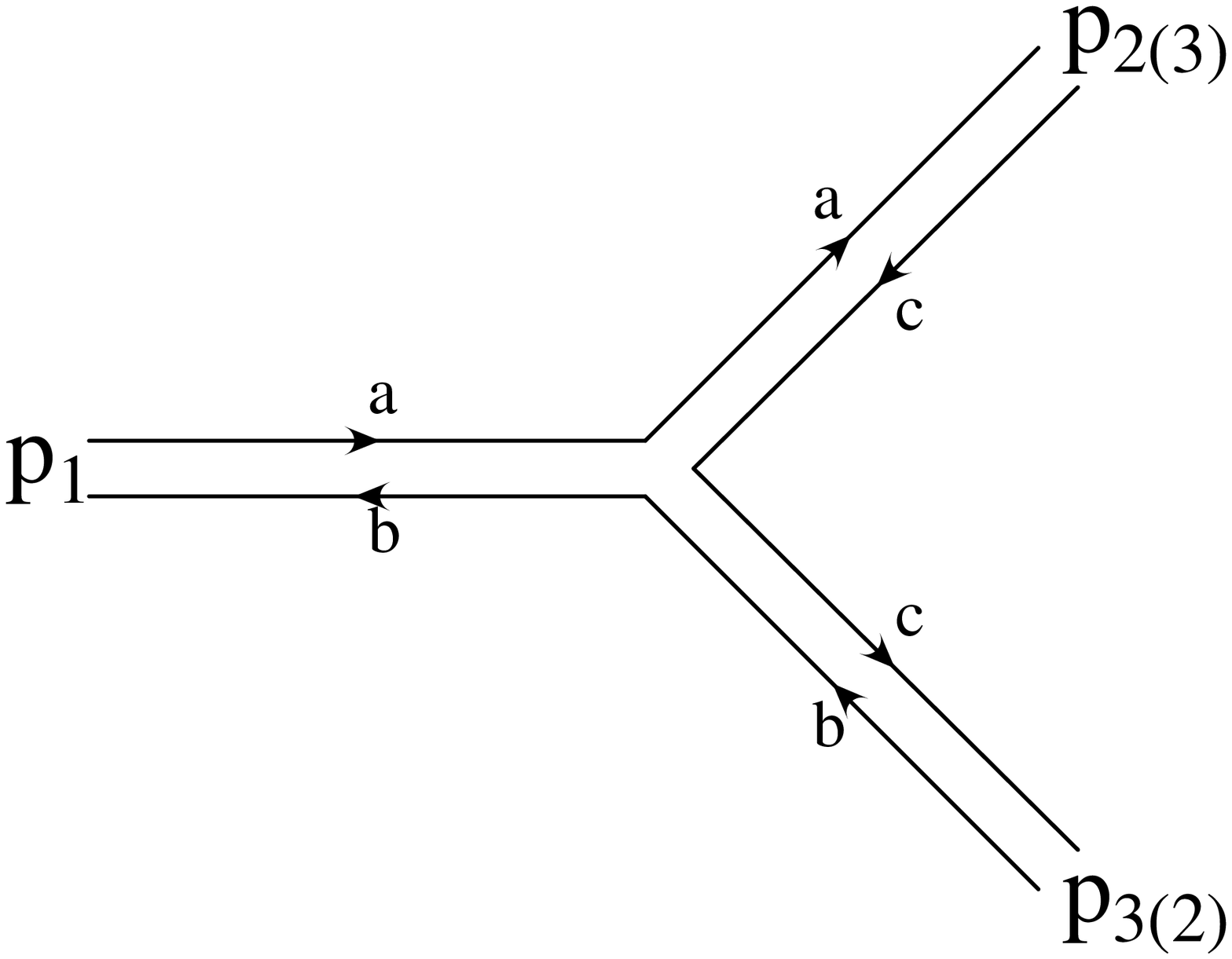}
\begin{center}  Figure 1.  \end{center}

\newpage
\hspace*{0.6cm} \epsfxsize = 13cm	\epsfbox{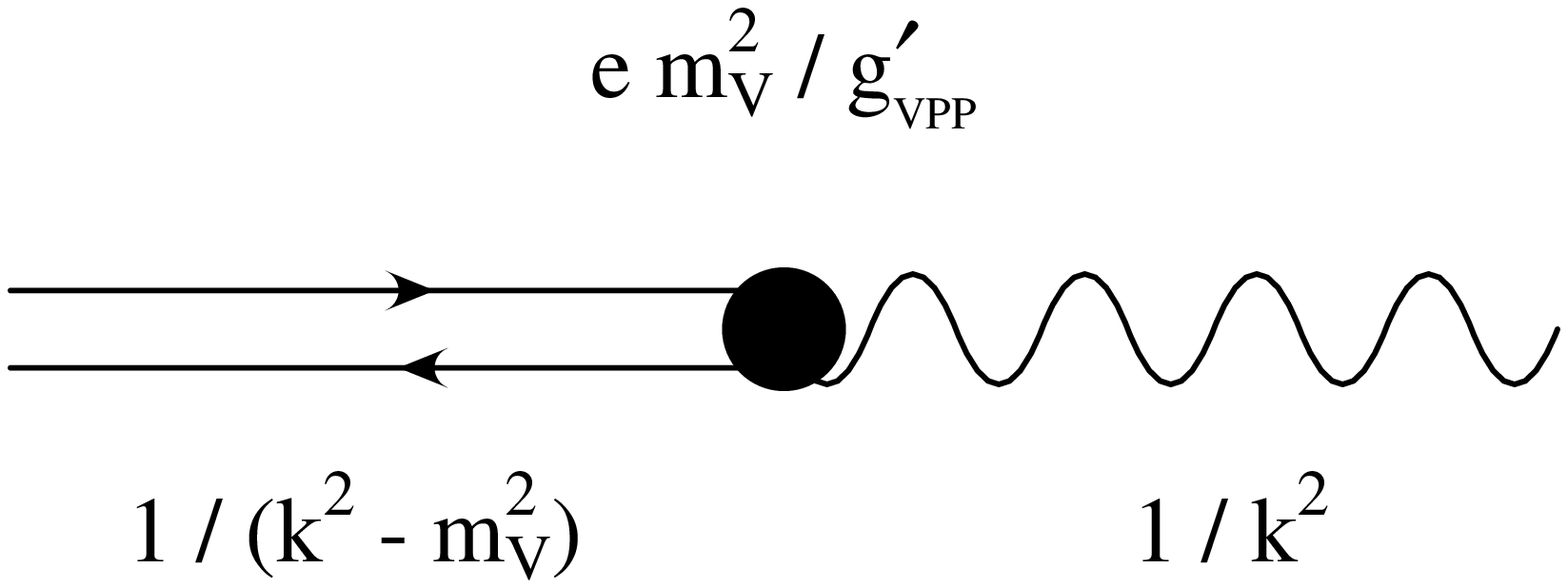}
\begin{center}  Figure 2.  \end{center}	

\newpage 
$\mo \gvvp - 2 \gvpp$\\
\hspace*{1cm} \epsfxsize = 11cm	\epsfbox{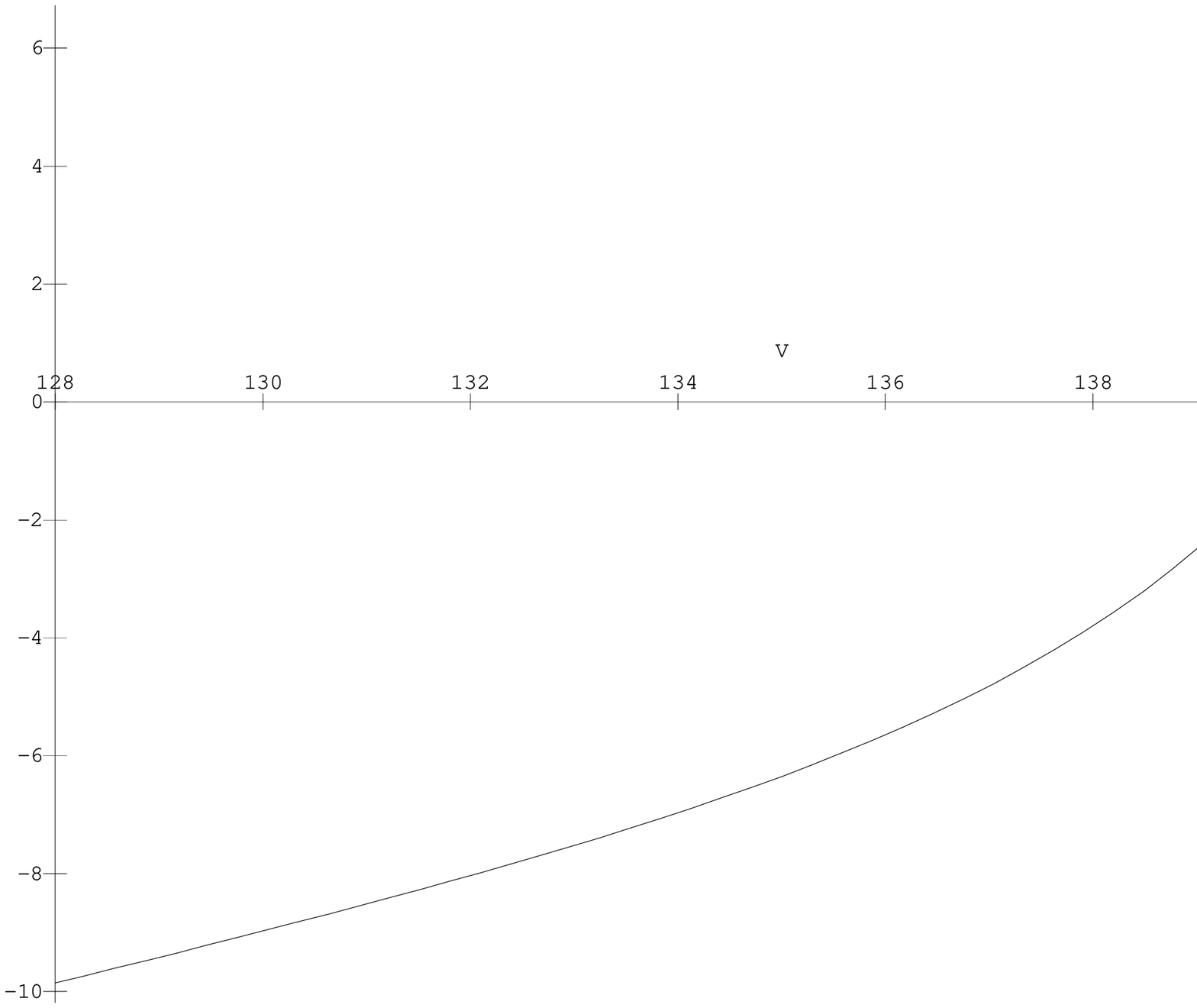}
\begin{center}  Figure 3.  \end{center}

\end{document}